\documentclass[aps,prd,preprintnumbers,showpacs,nofootinbib,twocolumn,10pt]{revtex4}
\usepackage{graphicx}
\usepackage{bm}
\usepackage{amsmath}
\usepackage{amssymb}
\usepackage{amsthm}
\usepackage{color}
\usepackage[usenames,dvipsnames]{xcolor}
\usepackage{latexsym}
\usepackage{epsfig}
\usepackage{textcomp}
\usepackage{hyperref}
\newcommand{\be}{\begin{equation}}
\newcommand{\ee}{\end{equation}}
\newcommand{\bea}{\begin{eqnarray}}
\newcommand{\eea}{\end{eqnarray}}
\newcommand{\bean}{\begin{eqnarray*}}
\newcommand{\eean}{\end{eqnarray*}}

\newcommand{\LCDM}{$\Lambda$CDM$\;$}

\newcommand{\lsim}   {\mathrel{\mathop{\kern 0pt \rlap
  {\raise.2ex\hbox{$<$}}}
  \lower.9ex\hbox{\kern-.190em $\sim$}}}
\newcommand{\gsim}   {\mathrel{\mathop{\kern 0pt \rlap
  {\raise.2ex\hbox{$>$}}}
  \lower.9ex\hbox{\kern-.190em $\sim$}}}

\newcommand{\jose}[1]{{\textcolor{black}{ #1}}}

\begin{document}

\title{Anisotropic expansion and SNIa: an open issue}

\author{{Jose} Beltr\'an Jim\'enez$^{a}$, {Vincenzo} Salzano$^{b}$ and {Ruth} Lazkoz$^{b}$}
\affiliation{${^a}$Centre for Cosmology, Particle Physics and Phenomenology,
Institute of Mathematics and Physics, Louvain University,
2 Chemin du Cyclotron, 1348 Louvain-la-Neuve (Belgium)\\}
\affiliation{$^b$Fisika Teorikoaren eta Zientziaren Historia Saila, Zientzia eta Teknologia Fakultatea, \\ Euskal
Herriko Unibertsitatea, 644 Posta Kutxatila, 48080 Bilbao.}

\date{\today}

\begin{abstract}
We review the appropriateness of using SNIa observations to detect potential signatures of anisotropic expansion in the Universe.
We focus on Union2 and SNLS3 SNIa datasets and use the hemispherical comparison method
to detect possible anisotropic features. Unlike some previous works
where non-diagonal elements of the covariance matrix were neglected,  we use the full covariance matrix of the SNIa data, thus obtaining more realistic and not underestimated errors. As a matter of fact, the significance of previously claimed detections of a preferred direction in the Union2 dataset completely disappears once we include the effects of using the full covariance matrix. Moreover, we also find that such a a preferred direction is aligned with the orthogonal direction of the SDSS observational plane and this suggests a clear indication that the SDSS subsample of the Union2 dataset introduces a significant biased, making the detected preferred direction  unphysical. We thus find that current SNIa surveys are inappropriate to test anisotropic features due to their highly non-homogeneous angular distribution in the sky. In addition, after removal of the highest inhomogeneous sub-samples, the number of SNIa is too low. Finally, we take advantage of the particular distribution of SNLS SNIa sub-sample in the SNLS3 data set, in which the observations were taken along 4 different directions. We fit each direction independently and find consistent results at the 1$\sigma$ level. Although the likelihoods peak at relatively different values of $\Omega_m$,
the low number of data along each direction gives rise to large errors so that the likelihoods are sufficiently broad as to overlap within 1$\sigma$.
\end{abstract}


\maketitle

\section{Introduction}

Recent results from the \textit{Planck} mission \citep{Planck} have strengthened the status of the $\Lambda$CDM model \citep{LCDM} as the standard cosmological
model driving the dynamics of
our Universe. This evidence should not be lightly assumed though, as the $\Lambda$CDM model implicitly requires the validity of a large number of assumptions:
the Cosmological Principle \citep{CosmoPrinciple}, i.e., that our Universe is homogeneous and isotropic on sufficiently large scales; the full validity of General Relativity
all the way up to the horizon scale; the existence of unknown dark matter to grow structures via gravitational collapse; the presence of an unnaturally small
cosmological constant, which drives the present acceleration of our Universe; and a nearly scale invariant gaussian primordial spectrum of perturbations
generated during an inflationary epoch in the early universe. Each of them might be individually questioned \citep{AgainstLCDM}.  Among them, we will focus
here on he validity of the Cosmological Principle and analyse the
possibility of detecting a certain amount of anisotropy by using SNIa observations.

Theoretically motivated models giving rise to a violation of the Cosmological Principle by inducing a late-time anisotropic expansion of the universe have been extensively
considered in the literature. In \cite{Campanelli:2006vb}, it was argued that the presence of large scale (homogeneous) magnetic fields \cite{Campanelli:2006vb} could induce
a certain level of eccentricity in the universe expansion that might even solve the low CMB quadrupole problem. The same kind of anisotropic expansion can be achieved by
assuming that the dark energy large scale rest frame might differ from that of radiation and/or matter, giving rise to a cosmology with moving fluids. Within this scenario,
the dipole acquires a cosmological contribution \cite{MovingDipole}, the CMB quadrupole is also modified because the relative motion introduces a certain level of eccentricity
\cite{Beltran Jimenez:2007ai} and large scale flows of matter are generated \cite{Jimenez:2008vs}. The possibility of having different large-scale frames for
dark matter and dark energy has also been considered more recently in \cite{Harko:2013wsa}. Another  scenario with anisotropic expansion was proposed in \cite{Rodrigues:2007ny},
where the effects of having an anisotropic cosmological constant was studied. More generally,  models with anisotropic dark energy
\cite{Koivisto:2007bp,Battye:2006mb,Campanelli:2011uc,Linder} or anisotropic curvature \cite{Koivisto:2010dr} have also been considered.

Most of the models  leading to anisotropic expansion discussed in the previous paragraph make the universe metric be of the Bianchi I type, that is,
the universe expands at different rates along the 3 spatial directions. Some of those models are characterized by the presence of a privileged direction,
which is identified with the vector characterizing the model, direction of the magnetic field, direction of the relative motion between different species.
In these cases, the metric is restricted to be of the axisymmetric Bianchi I type in which the expansion rate along the privileged axis differs from the expansion rate along the orthogonal directions.
Of course, the amount of anisotropy that these models can generate while being compatible with the highly isotropic CMB is tightly constrained. The main effect comes from the Integrated Sachs-Wolf effect, which is an accumulative effect from the last scattering surface until today. Thus, for models with a non-dynamical evolution of the eccentricity like models with anisotropic equation of state, such anisotropy is essentially constrained to be less than $\lsim 10^{-4}$ \cite{Appleby:2009za}. However, if the source of anisotropy is dynamical or there are compensating effects, the constraints are less stringent. In fact, having a period of anisotropic expansion at low redshift would mainly affect the CMB quadrupole, which is affected by a large cosmic variance so that it has less constraining power than the higher multipoles.

Observations of SNIa have been proposed and used as probes of large scale anisotropies. One of the first attempts to constrain the isotropy of the universe by resorting to SNIa measurements was made in \cite{Kolatt:2000yg}.  In \cite{Bonvin:2006en} the luminosity-distance dipole from the SNIa distribution was analysed and shown how to use it to obtain measurements of the Hubble expansion rate $H(z)$. Higher multipoles  and additional contributions to the luminosity distance were subsequently computed \cite{dLanisotropies}. SNIa measurements have also been used to study the isotropy of the Hubble diagram \cite{Schwarz:2007wf}, \jose{local bulk flows \cite{bulkflowsSN}}, the matter distribution \cite{Antoniou:2010gw} or \jose{the potential anisotropy of the deceleration parameter \cite{Cai:2011xs}}.  The possibility of constraining dark energy fluctuations by means of the luminosity distance was explored in \cite{Blomqvist:2010ky}. \jose{In \cite{Mariano:2012wx}, a dipole-like distribution for dark energy was
analyzed along with its possible correlation with the fine structure dipole.  Also a dipole-like distribution was considered in \cite{Cai:2013lja},  but applied to the luminosity distance with respect to the $\Lambda$CDM case}.  \jose{A fully Bayesian tool to search for systematic contamination in SnIa data was developed in \cite{Amendola:2012wc} and further extended, including searches for anisotropic signals, in \cite{Heneka:2013hka}. }

In the present work we will revisit the constraints that can be obtained on the anisotropy of the universe from SNIa observations. We shall mainly follow the same approach as in \cite{Antoniou:2010gw} where the authors used a certain version of the hemispherical comparison method to test the isotropy of the matter distribution at the background (homogeneous) level.  We intend to refine their approach by
including some influential subtleties concerning the SNIa observations. In particular, we show that including the full covariance matrix with the corresponding increase in the errors makes the significance of previously claimed preferred direction detections disappear. Moreover, we will show that such potential (although not statistically significant) preferred direction happens to be be aligned with the orthogonal direction to the plane of observation of the SDSS subsamples, being an indication of its biased origin due to the particular observation strategy (i.e., of its highly clustered angular distribution along only four directions in the sky) of such a subsample.

The paper is organized as follows:  in \S.~\ref{sec:data} we will describe the SNIa samples we have chosen for our analysis and their main properties; in \S.~\ref{sec:hemisphere} we will describe the original method we use to test the presence of anisotropic expansion and the main novelties we introduce; in \S.~\ref{sec:analysis_pre}  we apply
our method to a set of simulated data with a known anisotropic distribution in order to test its validity; in \S.\ref{sec:analysis_res} we finally show the results obtained from our analysis and in \S.\ref{sec:Conclusions} we conclude by discussing our results.

\section{SNIa data}
\label{sec:data}

This Section will be devoted to introducing the SNIa datasets that we will use later on for our analysis. We will discuss the $\chi^2$ estimator to be used when using SNIa as well as some  useful properties of our cosmological data sample which should be taken into account in order to obtain correct results.

The most updated SNIa samples so far are SNLS3 \cite{SNLS} provided by the \textit{SuperNova Legacy Survey} team and Union2 \cite{Union2} provided by the {\it SuperNova Cosmology Project} team. A newer collection of SNIa from the Union team is available and called Union2.1. It has $23$ more SNIa with respect to its predecessor. However,  we prefer to use the Union2 compilation in favour of making the comparison with older literature easier and more direct. Also, Union2.1 only adds $\approx 23$ SNIa, so we expect their statistical weight for our analysis to be not very significant. There is also a recent compilation with 112 new SNIa provided by the Pan-STARRS1 Medium Deep Survey team \cite{Rest:2013bya},  which we will not use because it has a non-homogeneous angular distribution (see Table.1 from \cite{Rest:2013bya}), which makes it unsuitable for testing anisotropy. It is worth stressing here nevertheless that we aim to testing the suitability of current SNIa datasets to seek for anisotropic features and we do not intend to obtain the most updated constraints on them (since, as we will show, current datasets are actually unsuitable for that purpose).

\subsection{Statistical background}

The $\chi^2$ estimator for SNIa observations is generally defined as
\begin{equation}
\chi^2 = \Delta \boldsymbol{\mathcal{F}} \; \cdot \; \mathbf{C}^{-1} \; \cdot \; \Delta  \boldsymbol{\mathcal{F}}, \;
\end{equation}
where $\Delta\boldsymbol{\mathcal{F}} = \mathcal{F}_{theo} - \mathcal{F}_{obs}$ is the difference between the observed and theoretical value of the observable quantity, $ \mathcal{F}$, and $ \mathbf{C}^{-1}$ is the inverse of the covariance matrix. For the SNLS3 compilation, the observable quantity  $ \mathcal{F}$ will be the SNIa magnitude $m_{mod}$ for SNLS3, defined by
\begin{equation}
\label{eq:m_snls3}
m_{\rm mod} = 5 \log_{10} [d_{L}(z; c_i) ] - \alpha (s-1) + \beta \mathcal{C} + \mathcal{M} \; .
\end{equation}
In this expression, the $c_i$ denote the set of cosmological parameters that are to be fitted, $\mathcal{M}$ is a nuisance parameter combining the Hubble constant $H_{0}$ and
the absolute magnitude of a fiducial SNIa and $d_{L}$ is the dimensionless luminosity distance:
\begin{equation}
\label{eq:dl_H}
d_{L}(z, c_i) = (1+z) \ \int_{0}^{z} \frac{\mathrm{d}z'}{E(z',c_i)} \; ,
\end{equation}
with $E(z)$  the dimensionless Hubble expansion function $H(z,c_i)/H_0$.

The SNLS3 team also provides the full multidimensional covariance matrix with statistical and systematic errors for all the physical quantities involved in their analysis, assuming $\alpha$ and $\beta$ as free fitting parameters:
\begin{equation}\label{eq:cov1}
\mathbf{\widehat{C}} = \sigma^2_{stat}\mathbf{\widehat{I}} + \mathbf{\widehat{v}_{0}} + \alpha^2 \mathbf{\widehat{v}_{a}}^2 + \beta^2 \mathbf{\widehat{v}_{b}}^2 + 2 \alpha \mathbf{\widehat{v}_{0a}} -2\beta \mathbf{\widehat{v}_{0b}} - 2\alpha \beta \mathbf{\widehat{v}_{ab}} \;
\end{equation}
with
\begin{eqnarray}\label{eq:cov2}
\sigma^2_{stat} &=& \sigma_{mB}^2 + \alpha^2 \sigma_{st}^2 + \beta^2 \sigma_{col}^2 + 2 \alpha \sigma_{ms} -2\beta \sigma_{mC} \nonumber \\
&-& 2\alpha \beta \sigma_{sC} + \sigma_{int}^2 + \sigma_{z}^2 + \sigma_{pec}^2
\end{eqnarray}
the diagonal elements of the statistical errors, which are respectively: errors on magnitude; errors on stretch; errors on  color; correlations between magnitude and stretch, magnitude and colors, stretch and colors; intrinsic dispersion errors; redshift errors; peculiar velocity errors; and: $\widehat{v}_{0}$ the out-of-diagonal statistical and systematic errors on magnitude; $\widehat{v}_{a}$ the same for the stretch; $\widehat{v}_{b}$ for the color; $\widehat{v}_{0a}$ for the correlation between magnitude and stretch; $\widehat{v}_{0b}$ for the correlation between magnitude and color; $\widehat{v}_{ab}$ for the correlation between stretch and color. In \citep{SNLS} it is also argued that a correlation between SNIa magnitude and the mass of the host galaxy might be present. To account for this effect, they propose to divide the total sample into two groups:
SNIa whose host galaxy mass is $< 10^{10}$ $M_{\odot}$ and SNIa with galaxy mass $> 10^{10}$ $M_{\odot}$. This division influences the definition of
the $\chi^2$; see the appendix of \citep{SNLS} and the two $\mathcal{M}$ values marginalization formulae that we will adopt for SNLS3.

On the other hand, the Union2 compilation observable is  the distance modulus $\mu$, defined by
\begin{equation}
\mu \equiv m_{\rm mod} - M = 5 \log_{10} [d_{L}(z, c_i) ] + \mu_{0} \; ,
\end{equation}
where $M$ is the absolute magnitude and $\mu_{0}$ is a nuisance parameter similar to SNLS3 parameter $\mathcal{M}$. The main difference with SNLS3 is that the $\alpha$ and $\beta$ parameters are fixed at a preliminary stage \citep{Union2} and the given full covariance matrix does not depend on them. Union2 files lack of the necessary data to distinguish the two host galaxy mass families, so that we will use the one-$\mathcal{M}$ value marginalization formulae in \citep{SNLS}.

\subsection{Angular distribution}

\begin{figure*}[ht!]
\centering
\includegraphics[width=8.4cm]{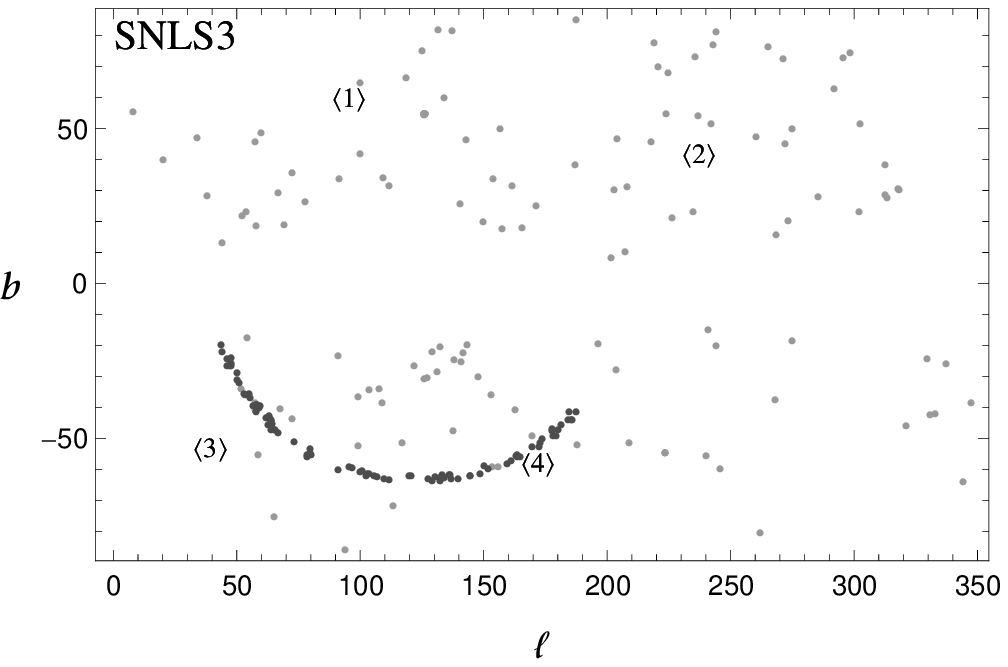}~~~
\includegraphics[width=8.4cm]{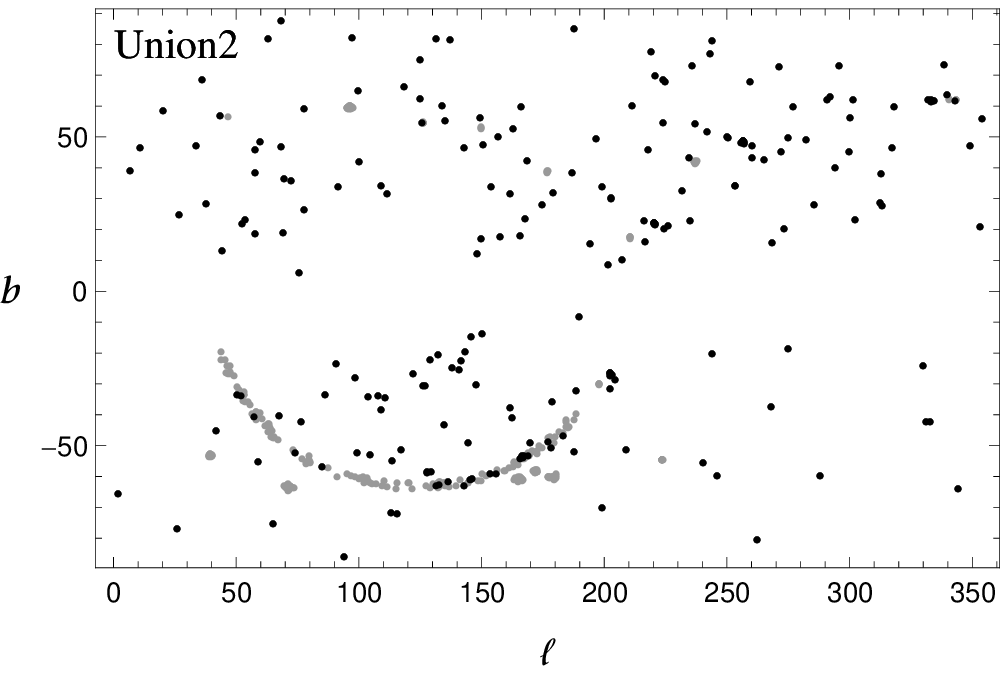} \\
\caption{SNIa sky distribution in galactic longitude $l$ and latitude $b$. \textit{Left panel.} SNLS3 SNIa sample: light grey points are Low-z SNIa; and Hubble data set; black points are SDSS SNIa; numbers in bracket are SNLS3 SNIa pointing toward different directions of  the sky. \textit{Right panel.} Union2 SNIa sample: light grey points are from SDSS, SNLS and smaller sub-samples with a not-homogeneous distribution in the sky; black points constitute the \textit{cut} Union2 sample and highlight sub-sample from the \textit{total} Union2 data set which show an homogeneous distribution in the sky.}\label{fig:SN_position}
\end{figure*}

Since the main aim of the present work is
testing the isotropy of SNIa samples, it is important to perform a preliminary analysis of the angular distribution of the samples that we are going to use.
This is  important since one could encounter cases with a false detection of a preferred direction in the universe, whose actual underlying reason might be a
non-homogeneous angular distribution of the sample. We also stress here that no tomographic analysis considering redshift distribution of SNIa will be performed in this work. Even if possible, it would reduce the number of SNIa data in each redshift bin, thus reducing the predictive power with too large errors on the cosmological parameters. We implicitly assume that the SNIa properties do not evolve throughout the expansion of the universe (at least for low redshifts where the SNIa are observed), but studying redshift-dependences of intrinsic SNIa properties is beyond the scope of the present work. We also comment here in advance that the main problem (discussed in next sections) will be related to the angular distribution of SNIa in the Union2 and SNLS3 datasets (mainly due to the fact that the SDSS sub-sample only gives data along 4 specific directions in the sky).

The SNLS3 sample is made out of four smaller sub-samples from four different surveys \cite{SNLS}: Low-z, \textit{SDSS}, \textit{SNLS} and SNIa from the Hubble Telescope. In the left panel of Fig.~\ref{fig:SN_position}
we show in galactic coordinates
the position of each SNIa in the sky , the longitude $l$ and the latitude $b$. SNIa from Low-z and Hubble samples are shown in light grey; SDSS SNIa are shown in black;
SNLS SNIa are identified by numbers in brackets. SDSS SNIa are distributed in the region
scanned by the SDSS survey and specifically chosen for the supernova survey project \cite{SDSS}. On the other side, SNLS SNIa show
a beam distribution: all ($242$) them are concentrated in four beams pointing toward different directions of the sky. Clearly,
this is a counter intuitive property which makes them unsuitable to be used in our analysis (this was also pointed out in
\citep{Schwarz:2007wf} where the authors referred to the very predecessor of this latest sample). However,
this sample is specially suited for the approach suggested in \cite{Linder}, since one can determine the cosmological parameters along
the four different directions independently. We will perform this analysis in \S.~\ref{sec:snls3}.

In the right panel of Fig.~\ref{fig:SN_position} we show the Union2 sample distribution in sky. It shares many objects in common with the SNLS3 sample, but with the addition of many smaller sub-samples. We still have SDSS and SNLS sub-samples in it, clearly identified in light grey circles.

Based on these considerations, and in order to develop our analysis, we will perform our statistical study considering two samples: the \textit{total} Union2 data set; and the so-called \textit{cut} Union2 one, which will correspond to the black points in Fig.~\ref{fig:SN_position}. The \textit{cut} sample is obtained considering only sub-samples which show an homogeneous distribution in space, so that, for example, SDSS, SNLS and others which have a peculiar angular distribution are not considered. The \textit{cut} sample will be made of $226$ SNIa, approximately the $40 \%$ of the \textit{total} sample. The cut-criterium has not been arbitrarily chosen but it is well-motivated by the results we will show in next sections, essentially consisting in removing the highly inhomogeneous subsamples.

Finally, we underline another important property of the SNIa distribution: the lack of observations in a narrow region about the galactic plane,
clearly identified by the condition $b=0$. Both SNLS3 and Union2 data sets show a void in the region $-15 \lesssim b \lesssim 15$; this is a consequence of the obscuration from excessive stellar density in such direction, and it is a problem that cannot be overcome in any way in this case.

\section{Hemispherical comparison method}
\label{sec:hemisphere}

This method was already used to look for asymmetries in the CMB in \cite{Eriksen:2003db}, where the hemispherical asymmetry anomaly was discovered. More recently, it has also been used
to look for evidences of anisotropic features in the Hubble diagram \cite{Schwarz:2007wf}, whereas a different version of it was used in \cite{Antoniou:2010gw}.
We will adopt the approach used in the latter.

The main idea behind it is to split the celestial sphere into two hemispheres and fit the corresponding cosmological data sets for each hemisphere independently.
The two hemispheres are determined by the corresponding {\it equatorial} plane identified, in our analysis, by the vector orthogonal to such a plane,
with galactic longitude $l_{a}$ and latitude $b_{a}$. Then, one looks for the splitting that yields the maximum difference for the cosmological parameters
of the best fit on each hemisphere.  It is maybe convenient to stress that this method is simply designed to look for anisotropic features within a given data set of cosmological observations, but it does not imply a direct link with the true universe model.

Obviously, this procedure will always give a \textit{maximum difference} direction, and one should study then the statistical
significance of such \textit{detection} as compared to the expected level of anisotropy in a pure isotropic model (that will always have some anisotropy due to statistical fluctuations).
In \cite{Antoniou:2010gw}, the authors fit the SNIa of each hemisphere to two \textit{independent} \LCDM models with vanishing
spatial curvature and two different matter density parameters.
Then, they search for the direction that maximizes the quantity
\be
\frac{\Delta \Omega_m}{\overline{\Omega}_{m}} \equiv 2\frac{\Omega_m^N-\Omega^S_m}{\Omega_m^N+\Omega^S_m}
\label{DOmega}
\ee
where $\Omega_m^{N,S}$ correspond to the values of $\Omega_m$ for the North and South hemispheres with respect to the chosen equatorial plane. This quantity can be interpreted as
a normalized difference between both values of $\Omega_m$. In \cite{Antoniou:2010gw}, the authors actually do not look for the maximum of this quantity. Instead, they generate a certain
number of random directions in the sky ($400$ directions for the Union2 data set, with about $280$ SNIa per hemisphere) and consider the maximum value of
$\frac{\Delta \Omega_m}{\overline{\Omega}_{m}}$
among the random distribution of generated directions. In this work we aim at revisiting this problem by making subtle methodological differences, which we describe and justify below, from which expectably more refined conclusions will emerge and which, why not, will perhaps eventually lead to different conclusions. Specifically, in this work we will consider the following points:
\begin{itemize}
   \item We will assume that each hemisphere is well-described by a $\Lambda$CDM model, although they might have a different matter density, $\Omega_{m}$. Despite the availability
   of more
   complex cosmological models involving a larger number of theoretical parameters, this choice seems a reasonably minimal extension and appropriate for our purposes.
   \item The main difference from previous works (specially with \cite{Antoniou:2010gw}) will be the use of Monte Carlo Markov Chain (MCMC) methods for the statistical analysis. After having split the data into two hemispheres, we will not select a certain amount of random directions but, instead, we will use an MCMC method to explore the entire space parameter and to obtain an angular map of the $\chi^2$ function.
   Then, we will be able to conclude which is the preferred direction of anisotropy of our SNIa sample, if any.
\end{itemize}
We will minimize the $\chi^{2}$ estimator using MCMC methods and testing their convergence with the power spectrum algorithm defined in \citep{Dunkley05}. In the case of the hemispherical comparison method, such an estimator is defined as explained in detail in the previous Section with the following Hubble expansion rate:
\begin{equation}
\label{eq:E_H}
E(z) = \left\{
        \begin{array}{ll}
           \left[ \Omega_{m}^{N} (1+z)^{3}+ (1-\Omega_{m}^{N}) \right]^{1/2}, & \text{if} \; \hat{r}\cdot\hat{r}_i>0  \\
           \left[ \Omega_{m}^{S} (1+z)^{3}+ (1-\Omega_{m}^{S}) \right]^{1/2}, & \text{if} \; \hat{r}\cdot\hat{r}_i<0
         \end{array}
       \right.
\end{equation}
where
\begin{eqnarray}
\hat{r} &=& (\cos l_{a} \cos b_{a}, \sin l_{a} \cos b_{a}, \sin b_{a}) \nonumber \\
\hat{r}_{i} &=& (\cos l_{i} \cos b_{i}, \sin l_{i} \cos b_{i}, \sin b_{i}) \; ,
\end{eqnarray}
are the directions identifying the equatorial plane and the directions of each SNIa in the sample, respectively, with $(l,b)$ the corresponding galactic coordinates. Of course, the $\Lambda$CDM model and spatial flatness are assumed in each hemisphere. This assumptions might be dropped and more general cases could be considered. The simplest step further could be the addition of an anisotropic dark energy equation of
state parameter $w$ \citep{Linder}, but this is out of the purpose of this work. Moreover, present data sets comprise too few SNIa already for the simplest case, so even poorer constraints will be obtained if we allow for additional free parameters.

It is worth to point out here that MCMC methods give us the possibility to easily implement priors on fitting parameters, if any physically well motivated reason is for them. In our analysis, we
have left the cosmological parameters $\Omega_{m}^{N}$ and $\Omega_{m}^{S}$ free to span the range $[0,1]$ with no other requirement. On the other hand, the equatorial planes coordinates
$l_{a}$ and $b_{a}$ can be constrained to a certain range due to evident symmetries considerations; we chose the ranges $0$\textdegree $\leq b_{a} \leq 90$\textdegree$\,$ and
$0$\textdegree $\leq l_{a} < 360$\textdegree. This last point has to be investigated in detail, as long as we do not know the degree of suitability
of MCMCs for such a test. The main parameters to be fit here will be the two cosmological values of $\Omega_{m}$, namely $\Omega_{m}^{N}$ and $\Omega_{m}^{S}$, and the two equatorial plane coordinates, i.e. $l_{a}$ and $b_{a}$. While it is well known the way $\Omega_{m}^{N}$ and $\Omega_{m}^{S}$ are related to observations and
theory, this is not the same for the equatorial plane coordinates. $\Omega_{m}^{N}$ and $\Omega_{m}^{S}$ are \textit{smoothly} varying quantities, in the sense that $\chi^2$ varies \textit{smoothly} when they are changed; on the other hand, a change in $l_{a}$ and $b_{a}$ might not be related to a change in $\chi^2$ for a given dataset. Depending on the spatial distribution of SNIa, there are voids in the SNIa distribution because of the few number of SNIa available for the whole sky, so that any value of $l_{a}$ and $b_{a}$ in that range would give the same result. In other words, for changes in  $(l_{a}, b_{a})$ such that the number of supernovae on each hemisphere remains the same, $\chi^2$ will not vary.

If we wanted to perform an analysis similar to \citep{Antoniou:2010gw}, we should use MCMC to minimize $\chi^{2}$ in each random direction as the authors do; but this would not bring any novelty to the topic. Here instead, we let the MCMC  explore freely all the parameter space, obtaining a full-sky angular $\chi^2$ distribution, from which we can derive the probability distributions of all the theoretical parameters involved. In other words, even if the MCMC will jump in a discrete way from one point to another, it will be the closest-to-continuous exploration method of the space parameter that can be realized in a non-hardware/time consuming effort. Notice that the MCMC will allow to refine the search as much as necessary.

As a further difference with previous works in the
literature, we will use the full statistical plus systematic non-diagonal covariance matrix error for SNIa. As we will show, this will  impact crucially the corresponding findings.
In \citep{Antoniou:2010gw} each hemisphere was independently fitted, i.e., the total $\chi^2$ was obtained as the sum of the two independent contributions obtained for each hemisphere $\chi^2_{\rm full\;sky}= \chi^2_{\rm north}+\chi^2_{\rm south}$. This approach was possible because the considered covariance matrix was diagonal, thus allowing the separation of the $\chi^2$ in the northern and southern terms so that minimization of each one can be independently performed. However, the use of only diagonal terms of statistical errors is well-known to produce a large underestimation of errors on cosmological parameters (up to $\approx 70\%$, see \citep{SNLS}) and might also produce a bias in the detection of an eventual anisotropy. Moreover, the independent fit of each hemisphere is not strictly correct, because another well-known effect is that SNIa are strongly correlated with each other thus resulting in a non-diagonal covariance matrix, which will play a crucial role for the significance of possible anisotropic features. Finally, in \citep{Antoniou:2010gw}, the parameter used to define the $1\sigma$ confidence
levels of the equatorial plane direction is strongly related to the errors on cosmological parameters (see their Eq.~(2.11)). It is thus clear that any larger contribution to such errors will produce larger errors for the equatorial plane coordinates, and so a less established detection of anisotropy. Having stablished the main lines of revision we continue to describe the setup still following the mentioned main references.

\section{Analysis: preliminary tests}
\label{sec:analysis_pre}

{\renewcommand{\tabcolsep}{1.mm}
{\renewcommand{\arraystretch}{1.5}
\begin{table*}[htbp!]
\begin{minipage}{\textwidth}
\caption{MCMC results.}\label{tab:results}
\centering
\resizebox*{0.75\textwidth}{!}{
\begin{tabular}{c|cc|cc}
\hline \hline
\multicolumn{5}{c}{mock I : known anisotropy vs homogeneous angular distribution} \\
\hline \hline
mock & $(\Omega_{m}^{N}-1\sigma,\Omega_{m}^{N}+1\sigma)$ &  $(\Omega_{m}^{S}-1\sigma,\Omega_{m}^{S}+1\sigma)$  & $(l_{a}-1\sigma,l_{a}+1\sigma)$ & $(b_{a}-1\sigma,b_{a}+1\sigma)$\\
\hline
\textit{total diag (I)} & $(0.520,0.586)$ & $(0.064,0.099)$ & $(133.81$\textdegree$,138.01$\textdegree) & $(26.43$\textdegree$,27.64$\textdegree) \\
\textit{total cov (I)}  & $(0.479,0.581)$ & $(0.048,0.098)$ & $(133.62$\textdegree$,137.51$\textdegree) & $(26.49$\textdegree$,27.62$\textdegree) \\
\textit{cut cov (I)}    & $(0.509,0.744)$ & $(0.041,0.154)$ & $(121.64$\textdegree$,147.97$\textdegree) & $(22.64$\textdegree$,39.95$\textdegree) \\
\hline \hline
\multicolumn{5}{c}{real data: unknown anisotropy vs real angular distribution} \\
\hline \hline
real & $(\Omega_{m}^{N}-1\sigma,\Omega_{m}^{N}+1\sigma)$ & $(\Omega_{m}^{S}-1\sigma,\Omega_{m}^{S}+1\sigma)$ & $(l_{a}-1\sigma,l_{a}+1\sigma)$ & $(b_{a}-1\sigma,b_{a}+1\sigma)$\\
\hline
\textit{tot diag} & $(0.239,0.287)$ & $(0.257,0.310)$ & $(0$\textdegree$,360$\textdegree) & $(0$\textdegree$,90$\textdegree) \\
\textit{tot cov}  & $(0.215,0.313)$ & $(0.251,0.354)$ & $(0$\textdegree$,360$\textdegree) & $(0$\textdegree$,90$\textdegree) \\
\textit{cut cov}  & $(0.208,0.302)$ & $(0.363,0.550)$ & $(132.16$\textdegree$,152.84$\textdegree) & $(26.39$\textdegree$,27.83$\textdegree) \\
\hline \hline
\end{tabular}}
\end{minipage}
\end{table*}}}

Prior to the application of the method to real data, we will first study and assess the following relevant aspects:
\begin{itemize}
   \item Suitability of MCMCs for anisotropy detection.
   \item Degradation in the anisotropy signal due to the use of the full covariance matrix error.
   \item Degradation in the anisotropy signal due to galactic plane.
\end{itemize}


The first point to be addressed is whether the use of MCMC is effectively suitable to look for a certain level of  anisotropy in the SNIa distribution.
For that, before using real SNIa data sets,
we will apply our algorithm to a set of simulated data with a homogeneous distribution in the sky data set of supernovae. We will endorse this simulated data with a known anisotropic
signal as follows:
We will demand a preferred direction determined by $\hat{r}$ and given by the equatorial plane galactic coordinates $l_{a} = 136.84$\textdegree$\,$ and $b_{a} = 27.07$\textdegree.
This plane came out as the preferential plane when the MCMC analysis was applied to the \textit{cut}  Union2 sample when the full covariance matrix error is used (see following sections).
Then, we generate a homogeneous set of random vectors $\hat{r}_i$ that will give the distribution of the SNIa in the sky. With this distribution, we confer each SNIa a distance modulus
following the $\Lambda$CDM model given in Eq.~(\ref{eq:E_H}), where we chose $\Omega_{m}^{N} = 0.5$ and $\Omega_{m}^{S} = 0.1$, and assuming that the Hubble constant is $H_{0} = 72$ km s$^{-1}$
Mpc$^{-1}$.  All the other
quantities needed, namely, redshifts and covariance matrix errors, are assumed to be exactly equal to those of the real Union2 dataset.

Another issue that we want to study is the following. Given that our galaxy prevents us from having observations of SNIa near the galactic plane (or, equivalently, a lack of data
points around that plane with respect to the rest of the sky) we want to
explore also whether the absence of SNIa data in the region near the galactic equator can introduce a potential bias or not. In particular, if the anisotropy happens
to be aligned with the galactic plane, this will certainly contribute to increase the errors in the position of the preferred axis. To study this effect we will use the mock SNIa
sample described  above and will remove all the SNIa in the range $-10$\textdegree$<b<10$\textdegree$\,$ around the galactic plane, and we will redistribute them outside that band.

Results are in Table~\ref{tab:results}. We compare three cases:
\begin{enumerate}
 \item the \textit{total} mock Union2 sample with homogeneous distribution and diagonal-only errors (named \textit{mock total diag (I)});
 \item the \textit{total} mock Union2 sample with homogeneous distribution and full covariance matrix (named \textit{mock total cov (I)});
 \item the \textit{cut} mock Union2 sample with homogeneous distribution, the galactic plane cut applied, and full covariance matrix (named \textit{mock cut cov (I)})
\end{enumerate}

We can point out the following results:
\begin{itemize}
   \item A homogeneously distributed sample is, of course, the perfect ideal case for a clear detection of an anisotropy signal.
   \item A more important point is that the MCMC method reveals to be fully suitable for our scopes: In all the cases considered
   the direction and the amount (different $\Omega_{m}$ values) of the anisotropy signal are quite clearly detected. We note, however, that even in this case there is an obvious degradation of the anisotropy signal, intrinsically due to the dispersion of data and to the observational errors. Indeed, such a degradation will make the detection in real data not very statistically significant.
   \item The use of diagonal-only errors or of the full covariance matrix influences only the amount of anisotropy, not its direction. We can easily see that the $1\sigma$ confidence interval for $\Omega^{N}_{m}$ and $\Omega^{S}_{m}$ is $\approx 65\%$ larger  than the
   diagonal-only errors case when the full covariance matrix errors is used.
   \item The direction of anisotropy is much more related to the number of SNIa included in the sample. This is clearly shown by comparing the \textit{total} ($557$ SNIa) and the \textit{cut} ($226$ SNIa) cases. We note that the \textit{cut} sample is derived assuming the lack of SNIa around the galactic plane too. Thus, less SNIa and the observational void around the galactic plane can: double the $1\sigma$ confidence interval of $\Omega^{N}_{m}$ and $\Omega^{S}_{m}$; and enlarge the $1\sigma$ confidence interval of equatorial plane coordinates, $l_{a}$ and $b_{a}$, of $\approx 10$ times.
\end{itemize}

\section{Analysis: results}
\label{sec:analysis_res}

Now that we have proved the suitability of the MCMC method by applying it to simulated data, we shall turn to real data. We shall first look at the Union2 compilation. We have applied the hemispherical comparison method and run a series of MCMC chains to obtain the best fit in three cases:
\begin{enumerate}
 \item the \textit{total} real Union2 sample with diagonal-only errors as in \citep{Antoniou:2010gw} (named \textit{real total diag});
 \item the \textit{total} real Union2 sample with full covariance matrix (named \textit{real total cov});
 \item the \textit{cut} real Union2 sample with full covariance matrix (named \textit{real cut cov});
\end{enumerate}

\begin{figure*}[htbp]
\centering
\includegraphics[width=7.5cm]{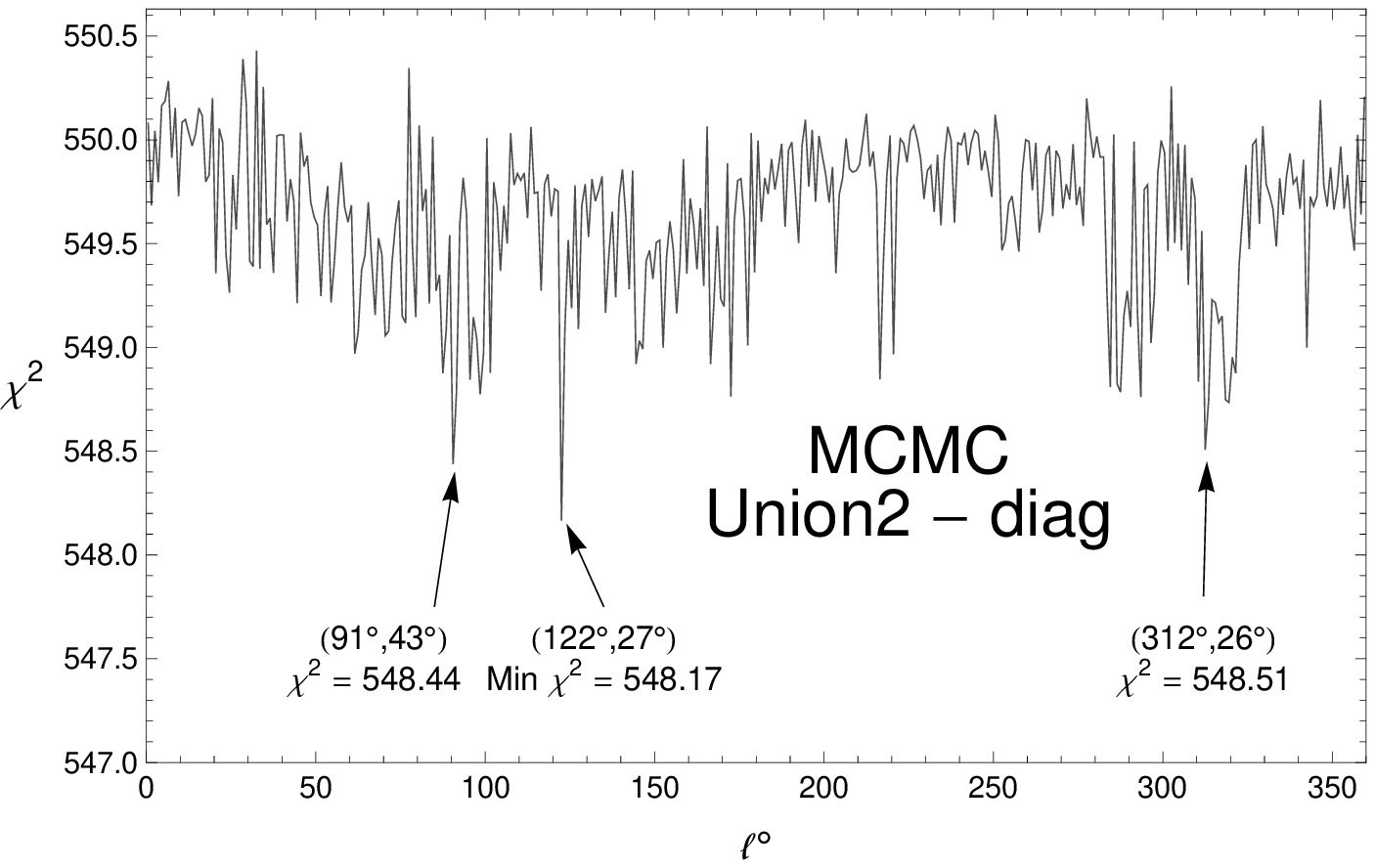}~~~~~~~~~~~~~~~~~~~~
\includegraphics[width=7.5cm]{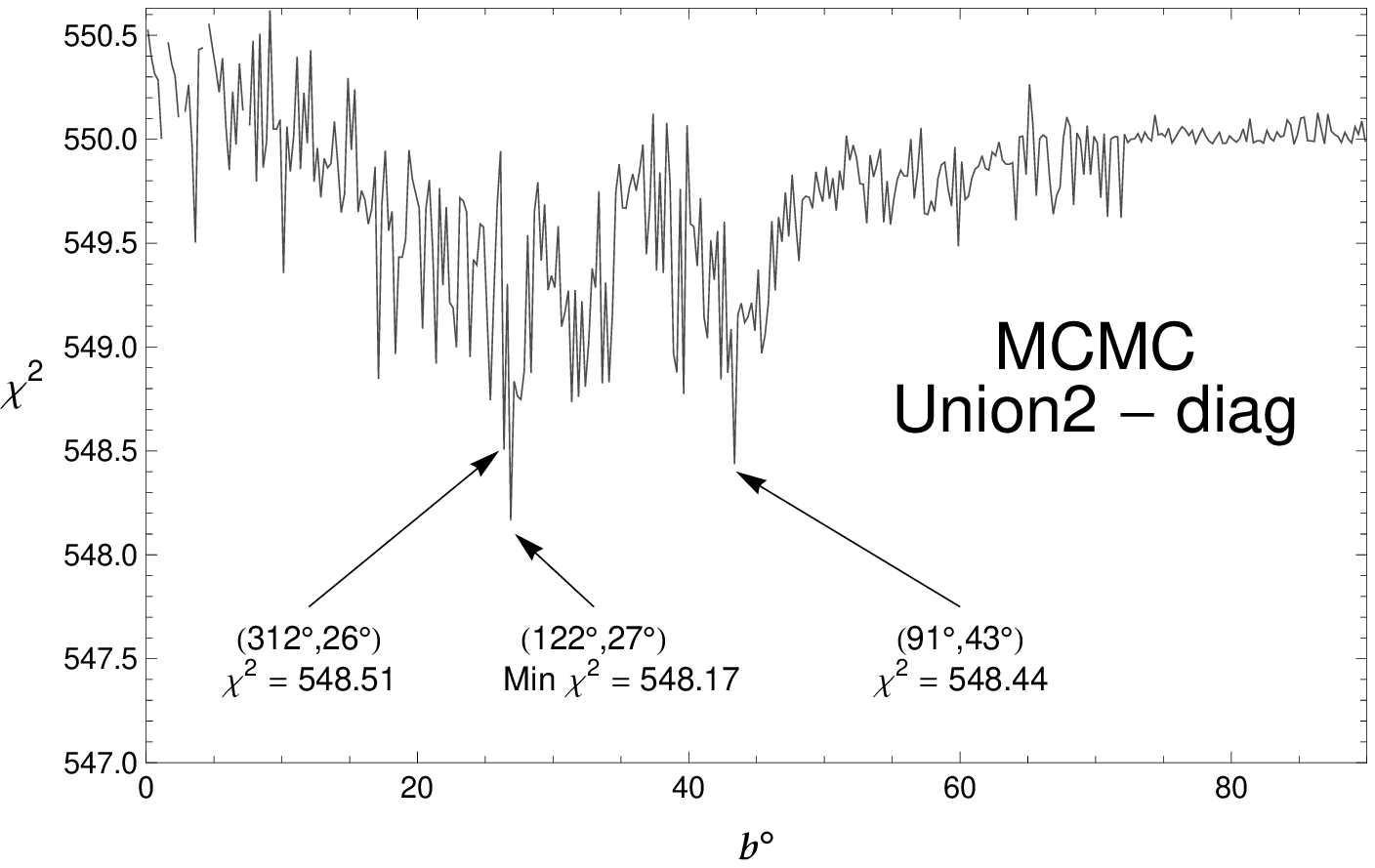}\\
~ \\
\includegraphics[width=7.5cm]{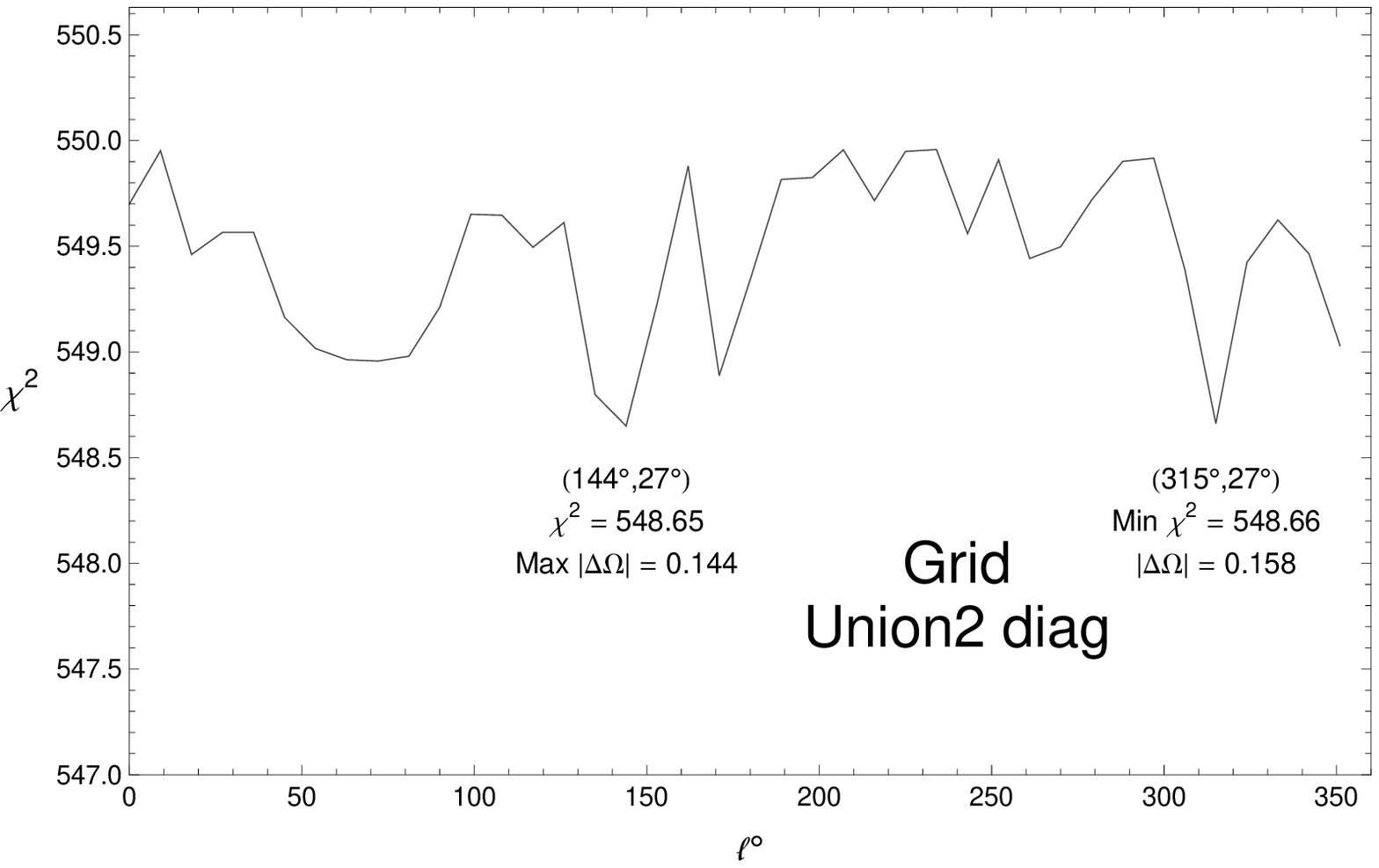}~~~~~~~~~~~~~~~~~~~~
\includegraphics[width=7.5cm]{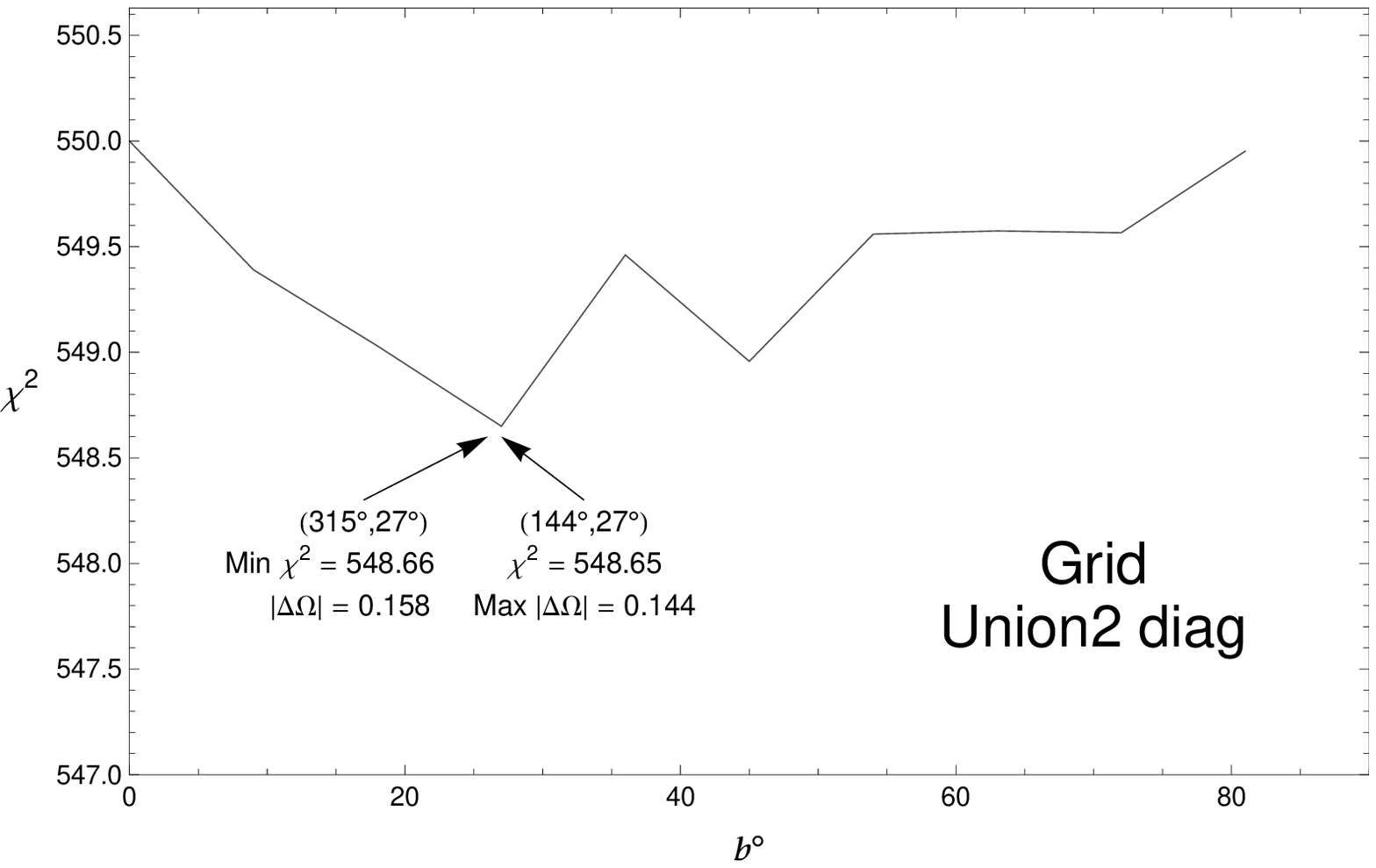}
\caption{Real Union2 with diagonal-only errors: projection of $\chi^2$ vs galactic longitude (left) and latitude (right) from MCMC (top) and regular grid (bottom).}\label{fig:comparison_diag_MCMC}
\end{figure*}

\begin{figure*}[htbp]
\centering
\includegraphics[width=7.5cm]{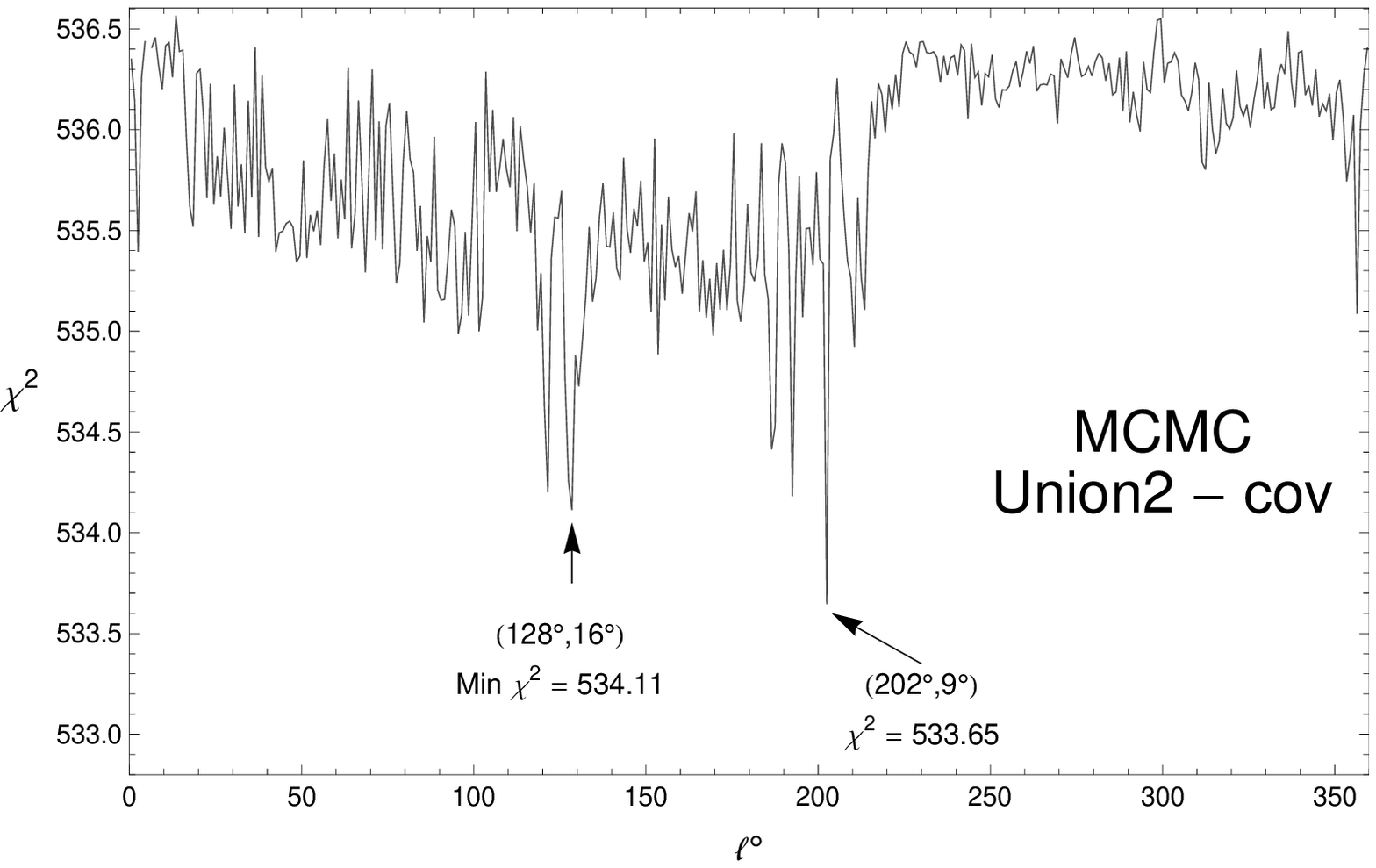}~~~~~~~~~~~~~~~~~~~~
\includegraphics[width=7.5cm]{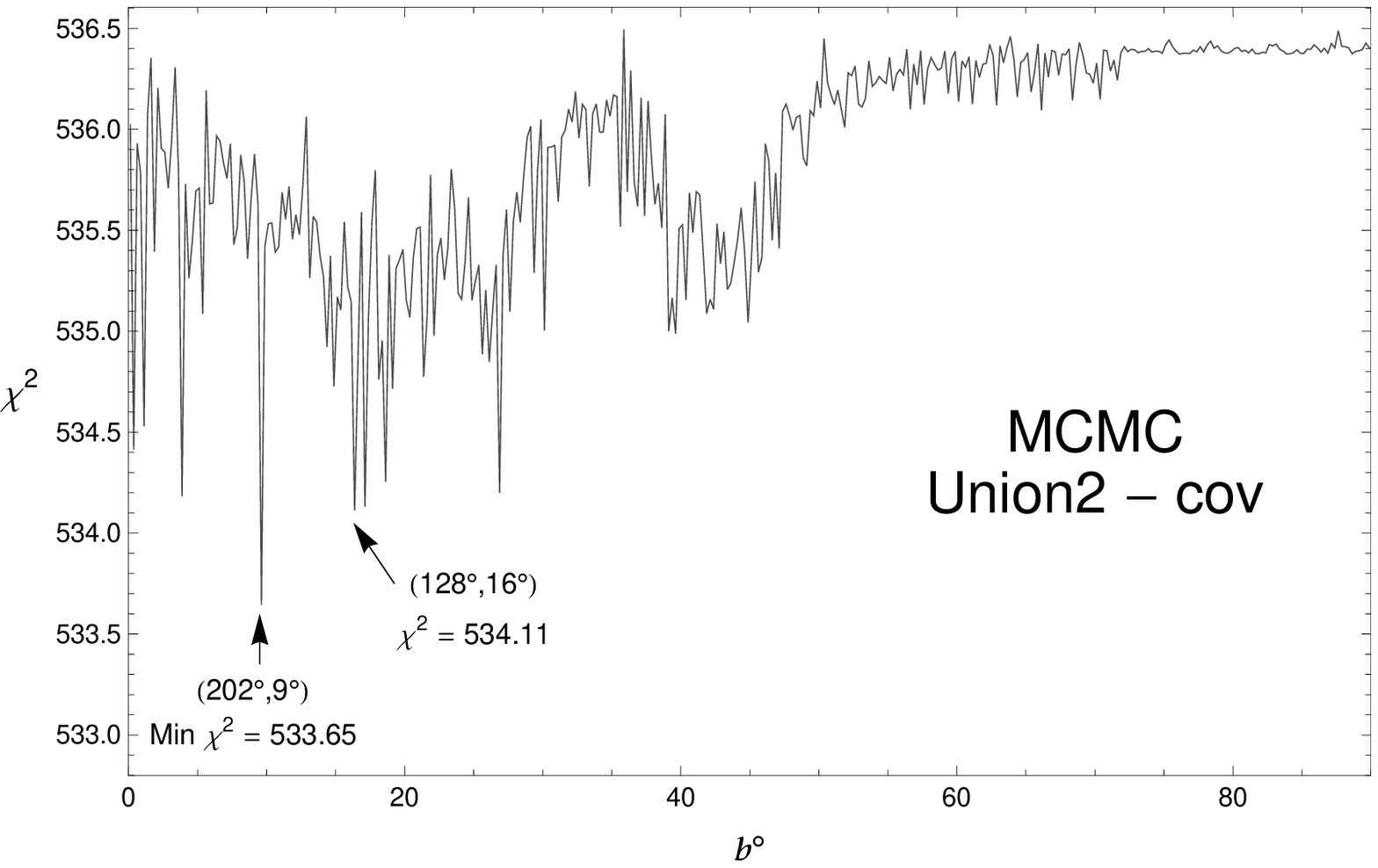}\\
~ \\
\includegraphics[width=7.5cm]{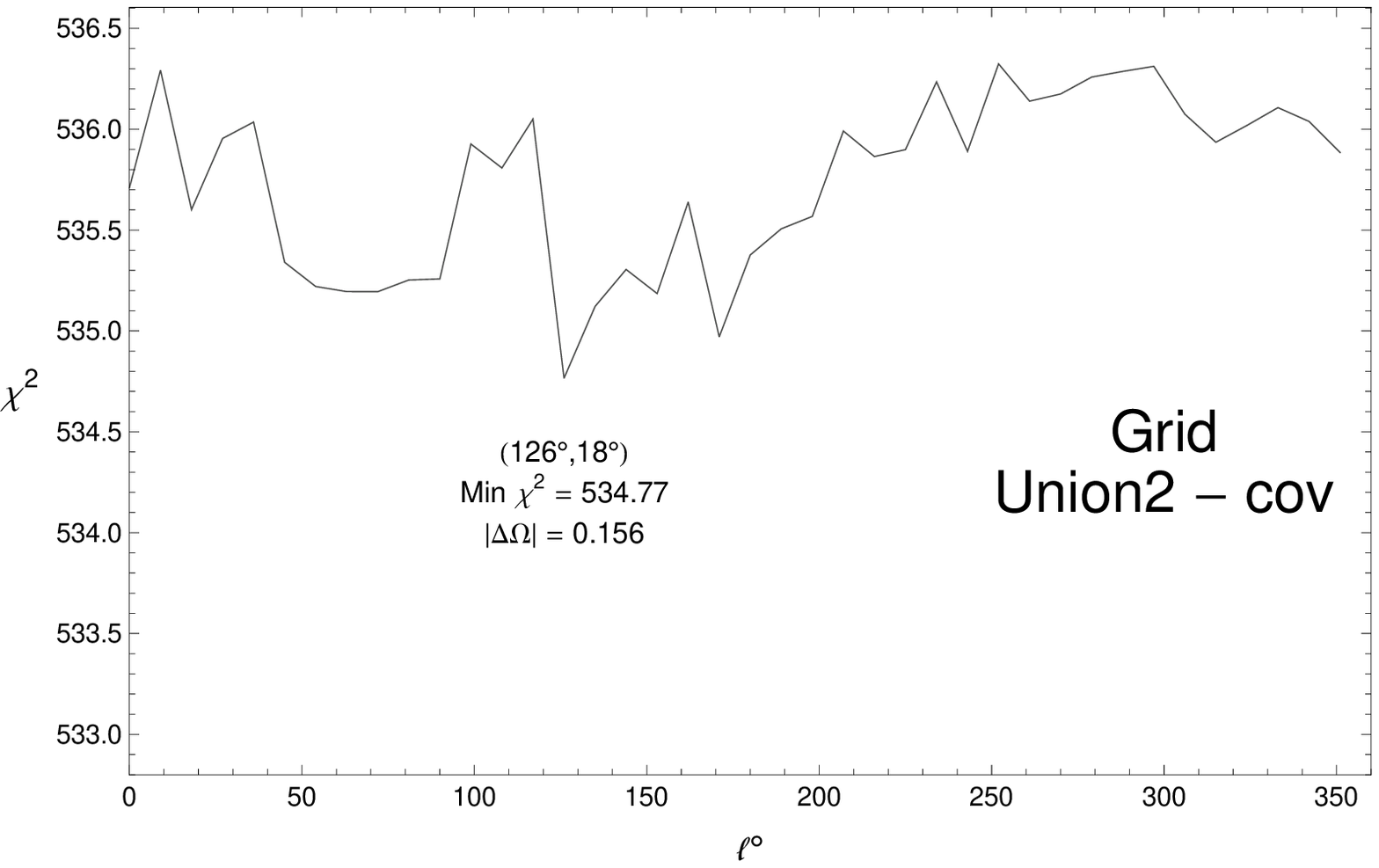}~~~~~~~~~~~~~~~~~~~~
\includegraphics[width=7.5cm]{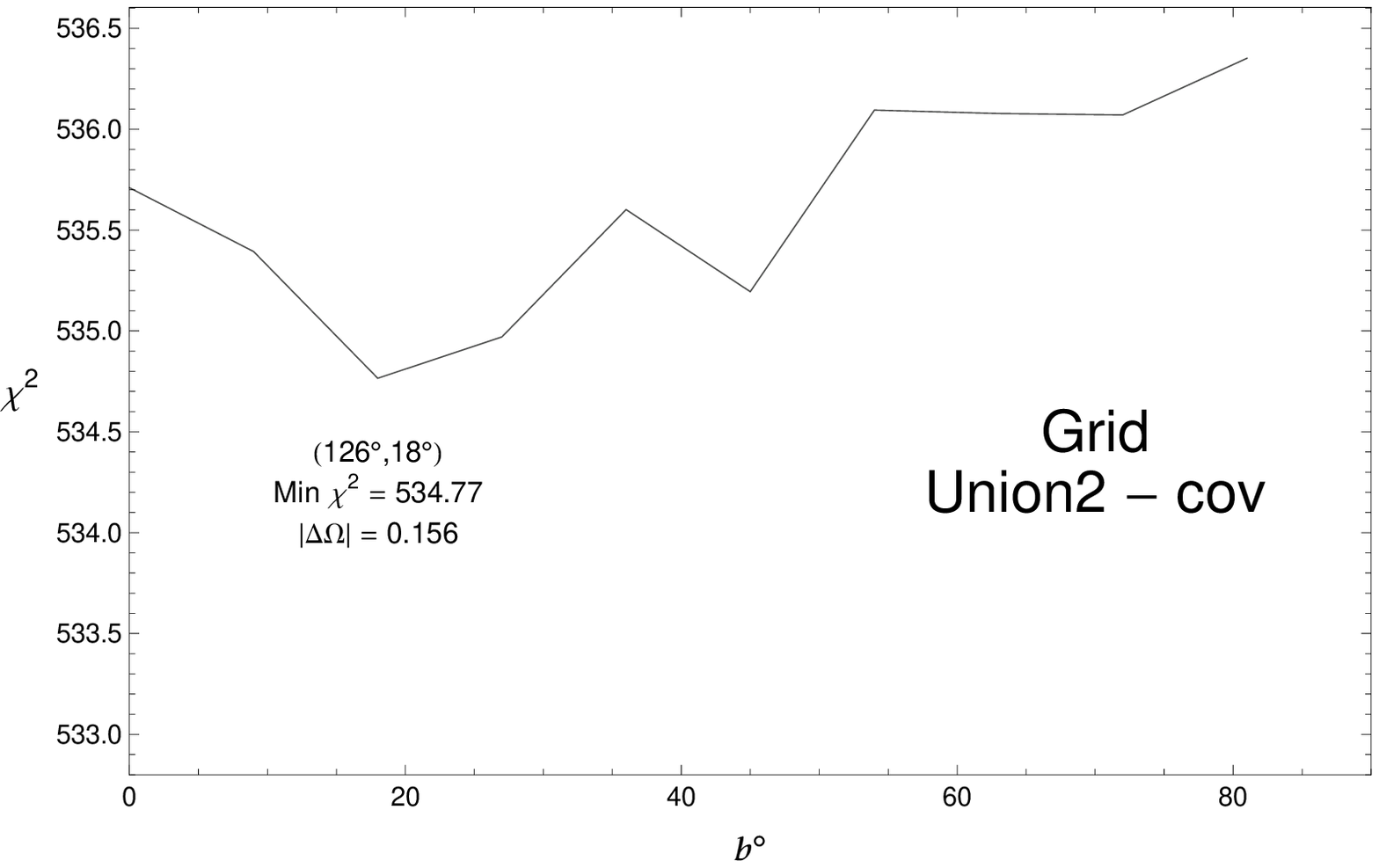}
\caption{Real Union2 with total covariance matrix: projection of $\chi^2$ vs galactic longitude (left) and latitude (right) from MCMC (top) and regular grid (bottom).}\label{fig:comparison_tot_MCMC}
\end{figure*}

\begin{figure*}[htbp]
\centering
\includegraphics[width=7.5cm]{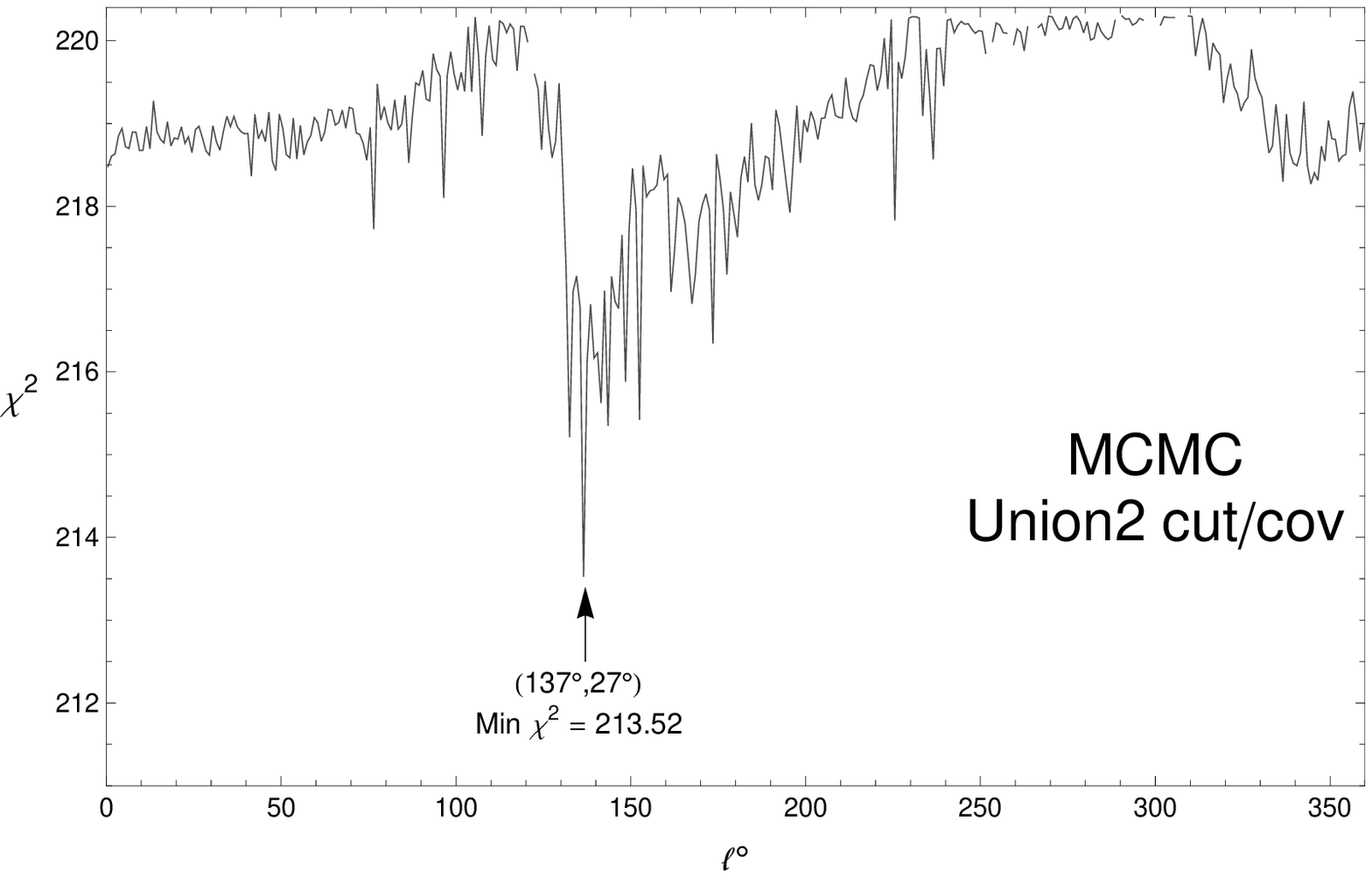}~~~~~~~~~~~~~~~~~~~~
\includegraphics[width=7.5cm]{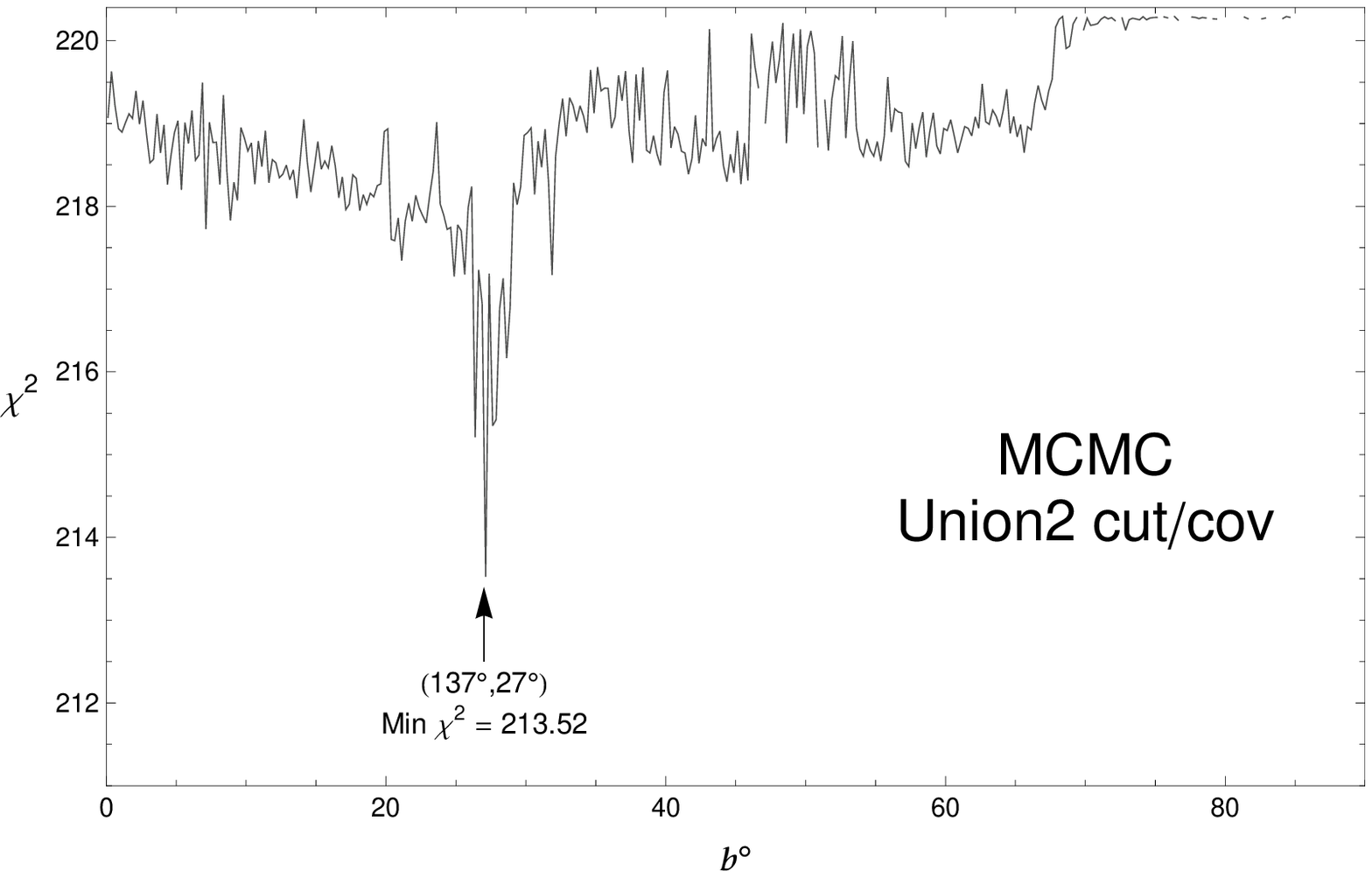}\\
~ \\
\includegraphics[width=7.5cm]{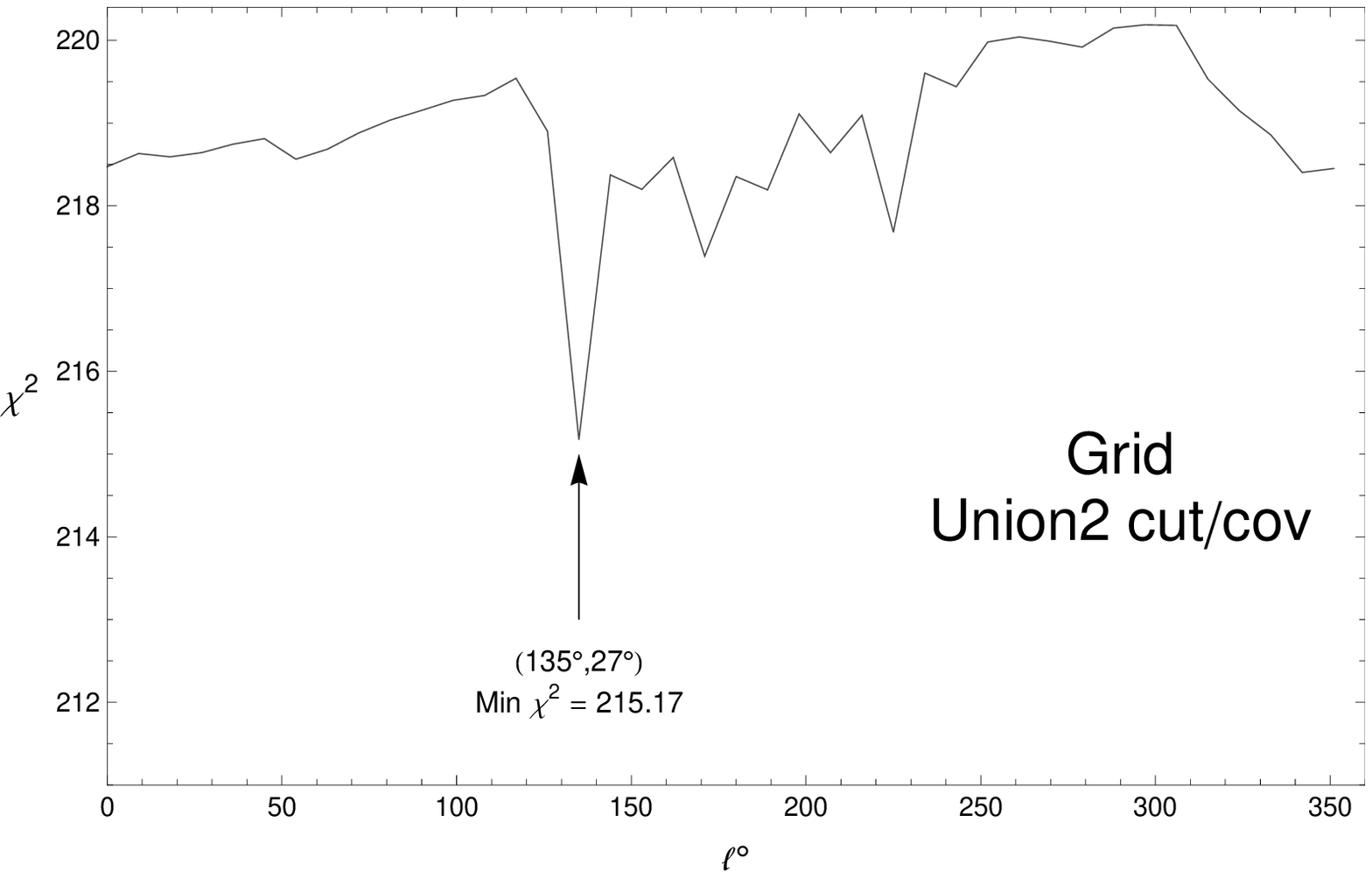}~~~~~~~~~~~~~~~~~~~~
\includegraphics[width=7.5cm]{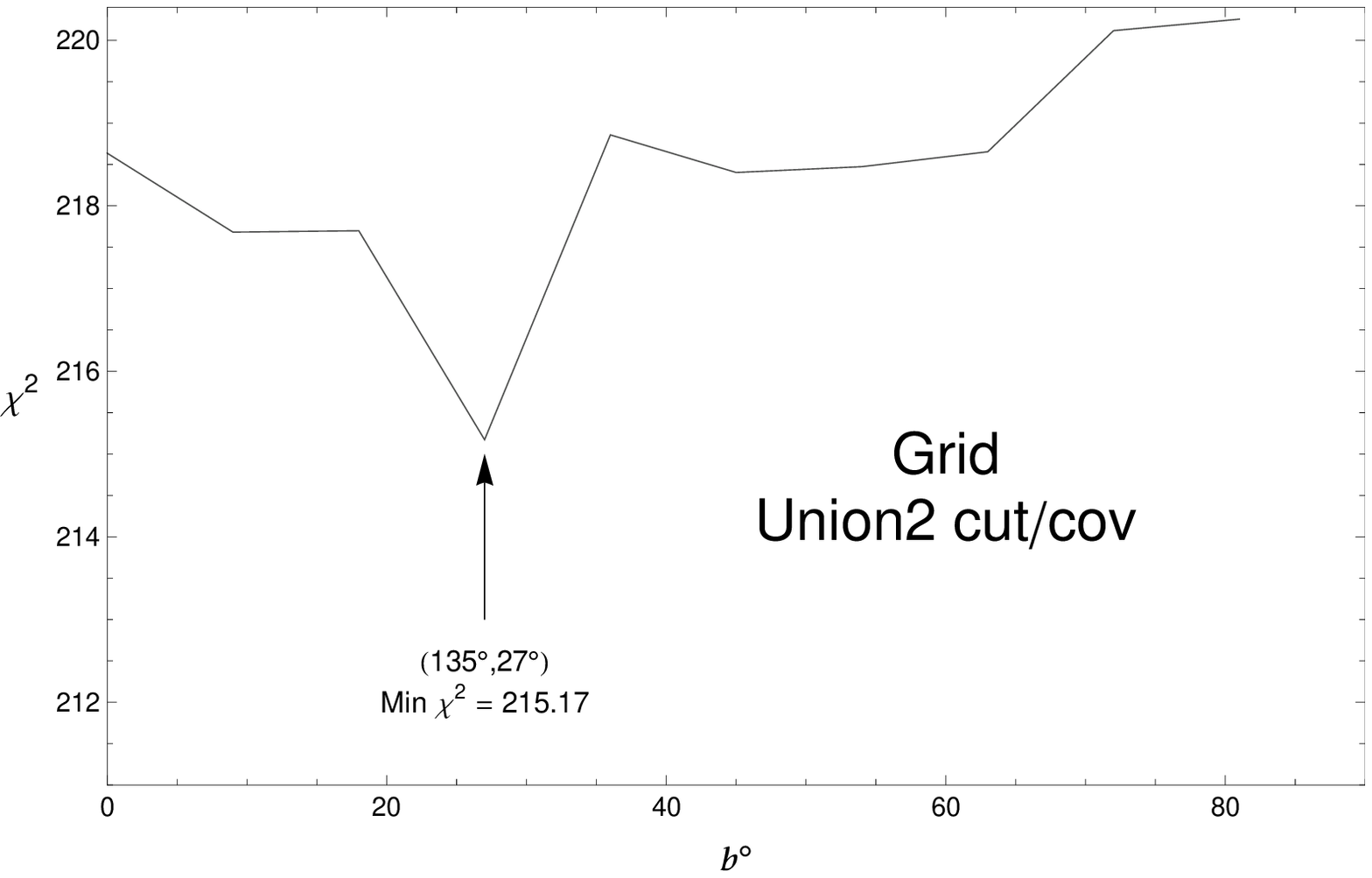}
\caption{Real cut Union2 with diagonal-only errors: projection of $\chi^2$ vs galactic longitude (left) and latitude (right) from MCMC (top) and regular grid (bottom).}\label{fig:comparison_cut_MCMC}
\end{figure*}

For each one of the previous cases, we have also performed an analysis similar to \citep{Antoniou:2010gw}: we have chosen $400$ directions, and found for the $\Omega^{N}_{m}$ and $\Omega^{S}_{m}$ values which minimize the $\chi^2$ function. As main differences with \citep{Antoniou:2010gw} we have:
\begin{itemize}
 \item built a regular grid of direction instead of choosing random directions. The coordinates have a grid step of $9$\textdegree$\,$ both in $l$ and $b$.  It would certainly be more appropriate to build a grid such that all sky cells have the same area. However, we have chosen a step-size such that we do not expect big differences between both ways of building the grid;
 \item minimized the \textit{total} $\chi^2$, instead of minimizing the north and south $\chi^2$ independently.
\end{itemize}
This has to be considered as a consistency test: by comparing MCMC results and the grid ones, we can test if and in what they differ, or not. Results are quite interesting and are shown in Figs.~\ref{fig:comparison_diag_MCMC}, \ref{fig:comparison_tot_MCMC} and \ref{fig:comparison_cut_MCMC}. We should remind here the main difference between the two methods. When using the grid, we minimize the $\chi^2$ with respect to $\Omega^{N}_{m}$ and $\Omega^{S}_{m}$ \textit{for each direction} and, consequently, in the corresponding figures we plot the \textit{best} $\chi^2$ values for each chosen direction (i.e. fixing $\Omega^{N}_{m}$ and $\Omega^{S}_{m}$ at the found best fit values). On the contrary, the MCMC shows us the \textit{full} angular dependence of the $\chi^2$ function, and the values in the figures could \textit{not} correspond to the the best $\chi^2$ values for each depicted coordinates set. In some sense, the information from MCMC is more complete and detailed because we perform a full
exploration of the spatial variation of the $\chi^2$.

\subsection{Total sample cases}

We can easily check by visual inspection, how MCMC reproduces at a very high accuracy the grid results, as a further confirmation of its goodness. With respect to the grid method, the MCMC have the improved feature of a complete span of the full parameter space. This feature eliminates the doubts about whether the number of grid/random directions is enough or not for the statistical analysis. In some cases we can also check one clear improvement from using MCMC: it finds some set of preferred parameters which cannot be analyzed by the grid method, neither with a regular grid or a random directions selection, as in \citep{Antoniou:2010gw}. Finally, it gives a direct and straightforward system to derive errors on all the parameters involved in the analysis,
as they can be extracted directly from the MCMC outputs.

Concerning the case of diagonal-only errors, Fig.~\ref{fig:comparison_diag_MCMC}, we can see how the direction corresponding to the minimum of $\chi^2$ is slightly different
from the one corresponding to the maximum anisotropy parameter $|\Delta \Omega_{m}|$: the former corresponding to the plane identified by coordinates $(144$\textdegree$,27$\textdegree$)$,
the latter to $(315$\textdegree$,27$\textdegree$)$. This last direction is very close to the one identified in \citep{Antoniou:2010gw} (differences arise only from the randomness versus
regular grid choices we have used), but it is \textit{not statistically} preferred, as its $\chi^2$  is not the best
found. Thus, assuming this as the anisotropy axis is doubtful. Moreover, when moving to the total covariance error matrix, Fig.~\ref{fig:comparison_tot_MCMC}, we see how this direction now disappears, while a new one, namely $(202$\textdegree$,9$\textdegree$)$, is present. The MCMC confirms these directions, giving also a more complete sketch
of the remaining parameter space.

\begin{figure}[ht]
\centering
\includegraphics[width=8.4cm]{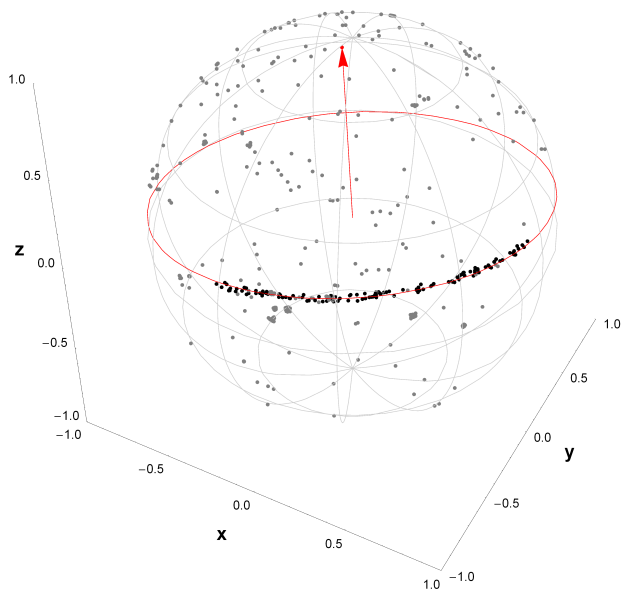}
\caption{This plot shows the distribution of SNIa from the Union2 compilation in the sky. The red circle and arrow represent respectively the anisotropy-equatorial plane and the orthogonal direction to such a plane with galactic coordinates $l_{a} = 122$\textdegree$\,$ and $b_{a}=27$\textdegree$\,$ from the \textit{real diag} Union2 analysis. Black points are SNIa from the SDSS sample; light gray points are SNIa from other sub-samples in the Union2 data set.}\label{fig:planes}
\end{figure}

The two cases, diagonal-only and full covariance matrix, also share a common property: in both of them, there is a direction which results to be associated with very low values of the $\chi^2$. It corresponds approximately to
$l \approx (122$\textdegree$-128$\textdegree$)$ and $b \approx (16$\textdegree$-27$\textdegree$)$. These values \textit{might} have to be considered with more attention because they correspond to the SDSS observational plane. We show it in Fig.~\ref{fig:planes}: the plane described above is in red, while the SDSS SNeIa are shown as black points. The correspondence between the two is strikingly evident, and it is only slightly weakened when using the total covariance matrix for the higher error budget considered. If we think that the SDSS are approximately the $23\%$ of the total sample, and are highly clustered, they might be considered as an intrinsic strong bias in this analysis.

Finally, by looking at Table~\ref{tab:results} and at the very close values of the $\chi^2$ shown in the Figs.~2 and 3, we can see that in the \textit{total diag} and the \textit{total cov} cases, no anisotropy is found, neither for what concerns its amount ($\Omega^{N}_{m}$ and $\Omega^{S}_{m}$ are perfectly consistent) nor for its direction (no preferred axis is found). This is not in conflict with results in \citep{Antoniou:2010gw} if we consider that errors on $\Omega_{m}$ are approximately double when moving from a diagonal to a full covariance matrix. In \citep{Antoniou:2010gw}, using diagonal errors, it is  argued that there is not a clear evidence for anisotropy. With a full covariance matrix we can here assess that no anisotropy is found at all.

\subsection{Cut sample cases}

There seems to be a clear evidence that the statistical analysis using the full Union2 sample is dominated by the SDSS subset, aligned with SDSS scanning direction, and that such a subset is introducing a strong bias in the best fits. This is the reason for defining the \textit{cut} sample, where SDSS
and SNLS SNeIa are removed in order to have a more homogeneous distribution of supernovae. This might help to assess more clearly for a detected anisotropy, even though
the cut in the number of data will inevitably degrade the signal (if any), as we have shown in the preliminary analysis with mock data.

\begin{figure}[ht]
\centering
\includegraphics[width=8.4cm]{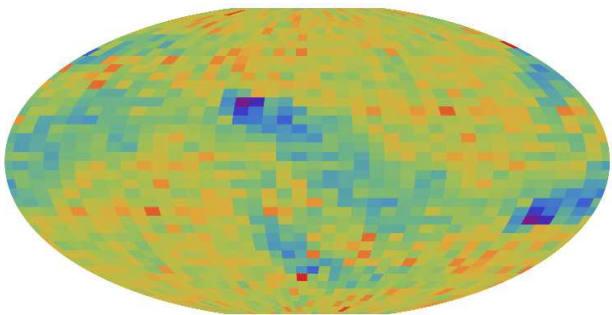}
\caption{Full-sky $\Delta \chi^2$ map. A grid with $800$ cells is considered; the spacing is $18$\textdegree in $l_{a}$ (horizontal axis) and $2.25$\textdegree in $b_{a}$ (vertical axis). Blue corresponds to $\Delta \chi^2 = 0$; red to $\Delta \chi^2 = 25$. }\label{fig:chi_map}
\end{figure}

Effectively, working with the \textit{cut} sample and the full covariance matrix gives much more interesting hints (see last row of Table.~\ref{tab:results}). In that case we found an anisotropic signal in terms of both amount and direction.
First, we wish to point out a quite clearly defined difference in the $\Omega_{m}$ values: the minimum in $\chi^2$ corresponds approximately to $\Omega_{m}^{N}\approx 0.25$ for the north pole, and $\Omega_{m}^{S} \approx 0.55$ for the south pole. 
Furthermore, we are also able to find an anisotropy direction, namely the orientation of the anisotropy-equatorial plane: $l_{a} = 137$\textdegree$\,$ and $b_{a} = 27$\textdegree. In Fig.~\ref{fig:chi_map} we show the $\chi^2$ full-sky distribution when a $\mathcal{N}=800$-cells grid is considered. Unfortunately, the low number of data left after the cut does not allow to give a high statistical significance to such a detection.

\begin{figure*}[ht!]
\centering
\includegraphics[width=8.4cm]{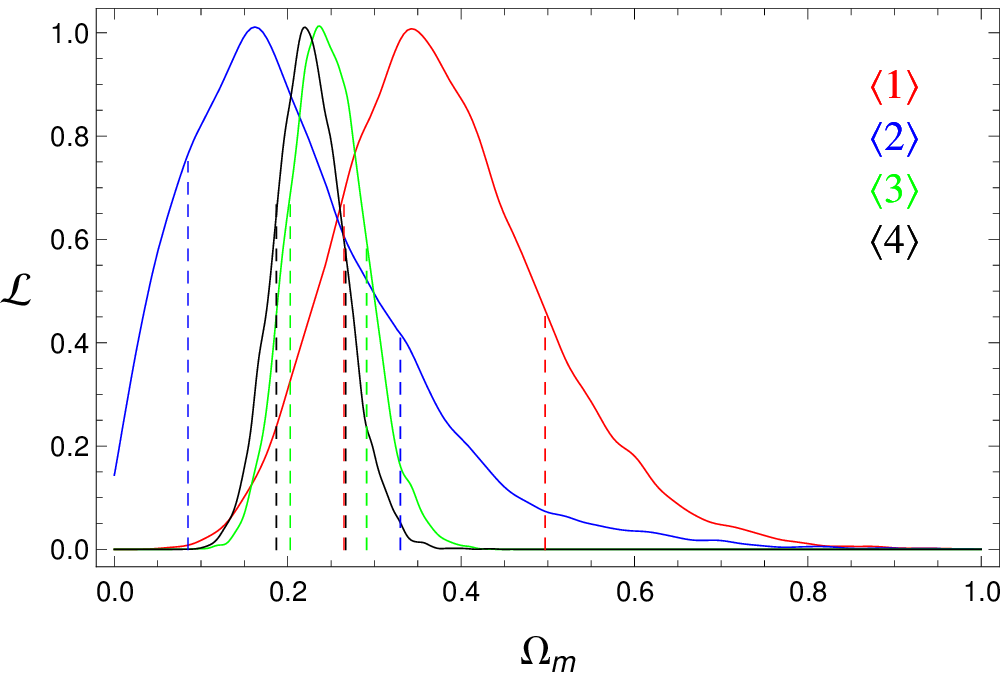}
\caption{Here we show the likelihoods for $\Omega_m$ as obtained from the 4 different beams of the SNLS3 compilation. We observe no significant tension between them.}\label{fig:SNLS3_beam}
\end{figure*}
\subsection{SNLS3 fits}
\label{sec:snls3}

Finally, we also consider the SNLS3 SNIa for our analysis. As we have discussed, their angular distribution can be seen as a paradigmatic badly suited sample  for testing isotropy on full-sky ranges. But their interesting feature of being aligned along four directions makes them a specially suitable dataset to look for deviations with respect to isotropy in a \textit{scanning} mode, in a way as the one suggested in \cite{Linder} to be the optimal manner. However, we have to keep in mind that the number of SNIa in each beam is still quite low and
a large error should be expected. In any case, this is a very interesting pattern that could be very useful for isotropy tests like the one performed here if a higher number
of supernovae is detected along each direction in future observations.

We have applied a line-of-sight approach to this compilation of SNIa by independently fitting the supernovae of each beam searching for a possible
direction-dependent change in $\Omega_{m}$. The results are shown in Fig.~\ref{fig:SNLS3_beam}:. We do not find any significant deviation from isotropy,
but only a small evidence for possibly different values of $\Omega_{m}$:
\begin{eqnarray}
\Omega_{m}^{<1>} &=& 0.368^{+0.129}_{-0.103} \nonumber \\
\Omega_{m}^{<2>} &=& 0.186^{+0.145}_{-0.101} \nonumber \\
\Omega_{m}^{<3>} &=& 0.245^{+0.046}_{-0.042} \nonumber \\
\Omega_{m}^{<4>} &=& 0.225^{+0.041}_{-0.038} \nonumber
\end{eqnarray}
The likelihoods of two of the beams (3,4) peak at $\Omega_m$ values which are very similar to each other, but the peak values for the other two are quite disimilar: one is rather
higher (1) and the other rather lower (2). However, the low number of SNIa ($\approx 60$ SNeIa in each beam)
makes the likelihoods be sufficiently wide as to overlap at the $1\sigma$ level. But it is not to be discarded that in the future, with more SNeIa, one could perhaps
be able to obtain stronger and more stringent constraints (or not).

\section{Conclusions}
\label{sec:Conclusions}
In this work we have presented some new updated results when searching for an anisotropy signal using SNIa measurements. We have used the hemispherical comparison
method consisting on fitting opposite hemispheres with independent $\Lambda$CDM models with the aim of finding an anisotropic distribution of $\Omega_m$. We have used
the Union2 compilation throughout our analysis. Unlike some previous studies, we have considered the full covariance matrix for the SNIa errors and found that it plays a very important role.
In fact, considering the full matrix introduces larger errors than those obtained when using only the diagonal errors. Thus, previously reported detections of anisotropy in SNIa observations
go away completely  when introducing the existing correlated errors. Moreover, we have used MCMC method, which implies a full detailed angular mapping of the $\chi^2$ function, allowing to discard possible false minima that could affect the grid or random orientations methods. Our result is in agreement with those in \cite{Heneka:2013hka}, where no significant indication of hemispherical anisotropy was found. Moreover, we have shown that the particular alignment of the SDSS SNIa seems to introduce a strong bias when searching for anisotropic features. On the other hand, we have used the SNLS3 data and have taken advantage of its very special distribution with the
SNIa lying along four different directions. We have then fitted each direction to an independent $\Lambda$CDM model and have obtained the
corresponding matter density parameters. Although the four directions yield likelihoods overlapping at the $1\sigma$ level, we have found that two of the directions give likelihoods that peak at quite different values of $\Omega_m$. However, the low number of SNIa in each direction does not allow to draw any significant conclusion.

\textbf{Acknowledgements}
J.B.J. is supported by the Wallonia-Brussels Federation grant ARC No.~11/15-040 and also thanks support from the Spanish MICINN Consolider-Ingenio 2010 Programme under grant
MultiDark CSD2009-00064 and project number FIS2011-23000. J.B.J. also wishes to acknowledges the Department of Theoretical physics of the University of the Basque Country
for their warm hospitality.  V.S. and R.L. are supported by the Spanish Ministry of Economy and
Competitiveness through research projects FIS2010-15492 and
Consolider EPI CSD2010-00064, and also by the Basque
Government through research project GIC12/66, and by
the University of the Basque Country UPV/EHU under
program UFI 11/55.

\end{document}